\documentclass{appolb}
\pdfoutput=1
\usepackage[pdftex]{graphicx}
\usepackage{amsmath}
\usepackage{braket}
\usepackage[T1]{fontenc}
\usepackage[english]{babel}
\usepackage{subfig}
\usepackage{float}
\usepackage{url}
\usepackage{braket}

\newcommand\T{\rule{0pt}{2.2ex}}
\newcommand\B{\rule[-1.0ex]{0pt}{0pt}}

\begin{document}

\title{Colour fields of the static hybrid gluon-quark-antiquark system
\thanks{Presented at the International Meeting "Excited QCD", Tatra National Park, Slovakia, 31 January - 6 February, 2010}
}
\author{Nuno Cardoso, Marco Cardoso and Pedro Bicudo
\address{CFTP, Instituto Superior Técnico\\ Avenida Rovisco Pais, 1, 1049-001 Lisboa, Portugal}
}

\maketitle
\begin{abstract}
The colour fields, created by a static gluon-quark-antiquark system, are computed in quenched SU(3) lattice QCD, in a $24^3\times 48$ lattice at $\beta=6.2$ and $a=0.07261(85)\,fm$. We study two geometries, one with a U shape and another with an L shape. The particular cases of the two gluon glueball and quark-antiquark are also studied, and the Casimir scaling is investigated in a microscopic perspective. This also contributes to understand confinement with flux tubes and to discriminate between the models of fundamental versus adjoint confining strings, analogous to type-II and type-I superconductivity.
\end{abstract}

\section{Introduction}
In this paper, we present a value for the the dual gluon mass in a SU(3) lattice QCD gauge independent and a detailed study of the Casimir scaling.
In section II, we introduce the lattice QCD formulation.
We briefly review the Wilson loop for this system and show how we compute the colour fields and as well as the lagrangian and energy density distribution.
In section III, the numerical results are shown.
We present results for the colour field profiles in the mid flux tube section for the static hybrid $gq\overline{q}$, in a U shape geometry.
A detailed study of the Casimir scaling is done and we present a value for the effective dual gluon mass and some values found in literature for the effective dual gluon mass and gluon mass.
Finally, we present the conclusion in section IV.

\section{The Wilson Loops and Colour Fields}

\begin{figure}[h]
\begin{centering}
    \subfloat[\label{loop0}]{
\begin{centering}
    \includegraphics[width=3cm]{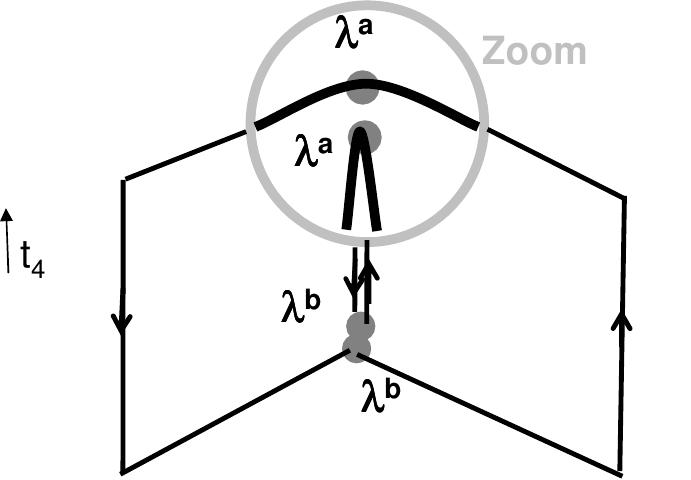}
\par\end{centering}}
    \subfloat[\label{loop1}]{
\begin{centering}
    \includegraphics[width=4cm]{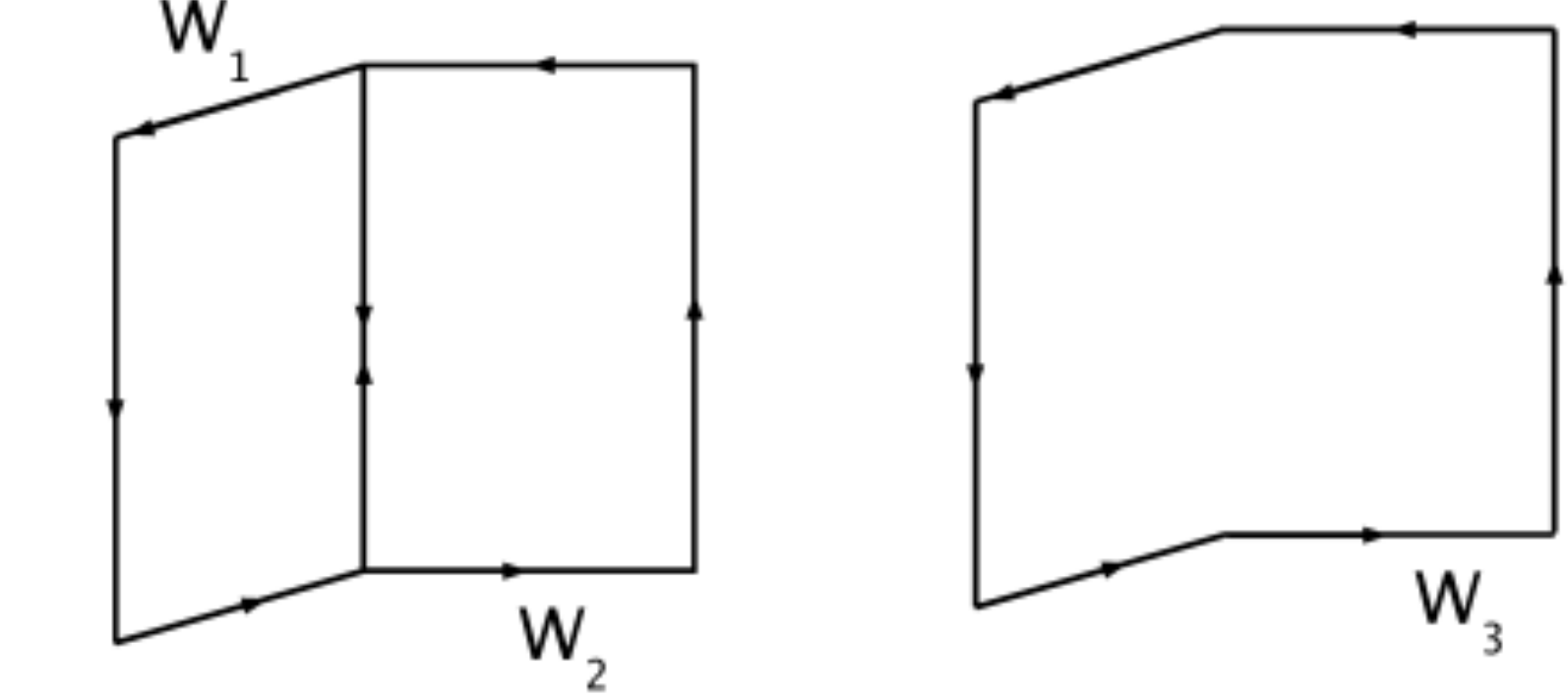}
\par\end{centering}}
\par\end{centering}
    \caption{\protect\subref{loop0} Wilson loop for the $gq\overline{q}$ and equivalent position of the static antiquark, gluon, and quark. \protect\subref{loop1} Simple Wilson loops that make the $gq\overline{q}$ Wilson loop.}
    \label{loop}
\end{figure}

The Wilson loop for the static hybrid $gq\overline{q}$ was deducted in \cite{Bicudo:2007xp,Cardoso:2007dc,Cardoso:2009qt,Cardoso:2009kz}, therefore we only present the fundamental expressions. The Wilson loop for this system is given by
\begin{equation}
    W_{gq\overline{q}}=W_1 W_2 - \frac{1}{3}W_3
\label{wloopqqg}
\end{equation}
where $W_1$, $W_2$ and $W_3$ are the simple Wilson loops shown in Fig. \ref{loop}.

In order to improve the signal to noise ratio of the Wilson loop, we use APE smearing, \cite{Cardoso:2009kz}, with $w = 0.2$ and iterate this procedure 25 times in the spatial direction. To achieve better accuracy in the flux tube, we apply the  hypercubic blocking (HYP) in time direction, \cite{Hasenfratz:2001hp}, with $\alpha_1=0.75$, $\alpha_2=0.6$ and $\alpha_3=0.3$. Notice that we only apply the smearing technique to the Wilson loop.

The chromoelectric and chromomagnetic fields are given by,
\begin{eqnarray}
    \Braket{E^2_i} & = & \Braket{P_{0i}}-\frac{\Braket{W\,P_{0i}}}{\Braket{W}}\\
    \Braket{B^2_i} & = & \frac{\Braket{W\,P_{jk}}}{\Braket{W}}-\Braket{P_{jk}}
\end{eqnarray}
where the $jk$ indices of the plaquette, $P$, complement the index $i$ of the chromomagnetic field.
The energy ($\mathcal{H}$) and lagrangian ($\mathcal{L}$) densities are given by
\begin{eqnarray}
    \mathcal{H} & = & \frac{1}{2}\left( \Braket{E^2} + \Braket{B^2}\right)\\
    \mathcal{L} & = & \frac{1}{2}\left( \Braket{E^2} - \Braket{B^2}\right)\,.
\end{eqnarray}

\section{Results}

Here we present the results of our simulations with 286 $24^3 \times 48$, $\beta = 6.2$ quenched configurations generated with the version 6 of the MILC code \cite{MILC}, via a combination of Cabbibo-Mariani and overrelaxed updates.
The results are presented in lattice spacing units of $a$, with $a=0.07261(85)\,fm$ or $a^{-1}=2718\,\pm\, 32\,MeV$.

\begin{figure}[h]
\begin{centering}
    \subfloat[U shape geometry.\label{fig:shapeU}]{
\begin{centering}
    \includegraphics[height=2.2cm]{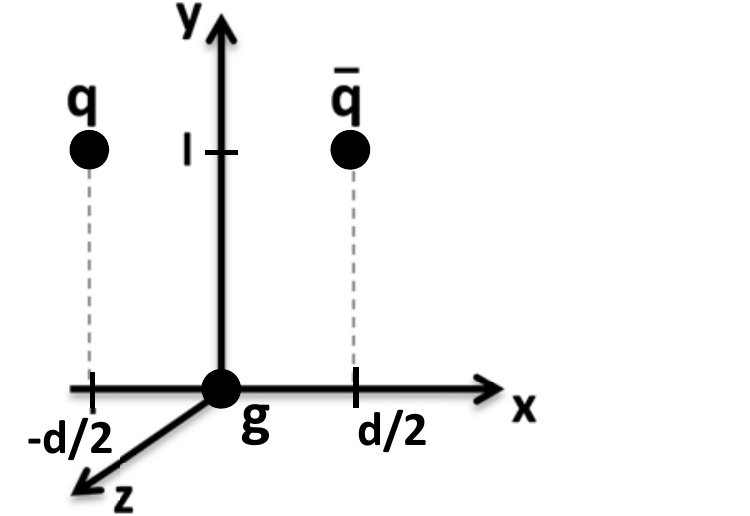}
\par\end{centering}}
    \subfloat[L shape geometry.\label{fig:shapeL}]{
\begin{centering}
    \includegraphics[height=2.2cm]{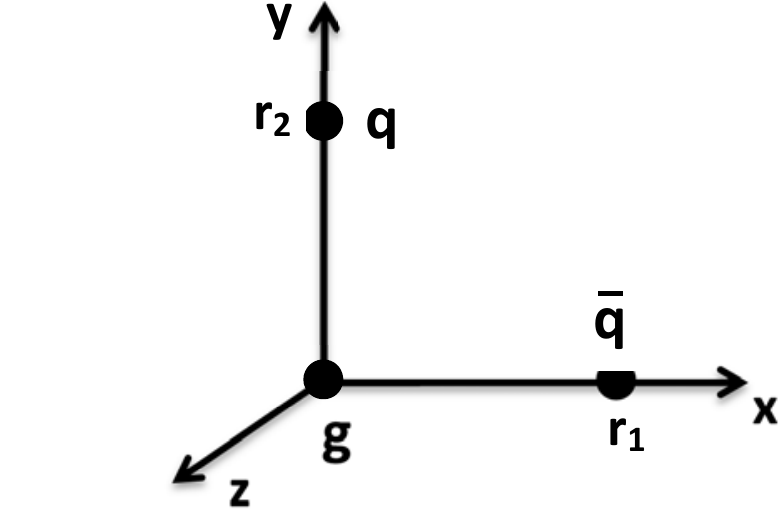}
\par\end{centering}}
\par\end{centering}
    \caption{gluon-quark-antiquark geometries, U and L shapes.}
    \label{shape}
\end{figure}

\begin{figure}[h]
\begin{center}
    \includegraphics[width=5.5cm]{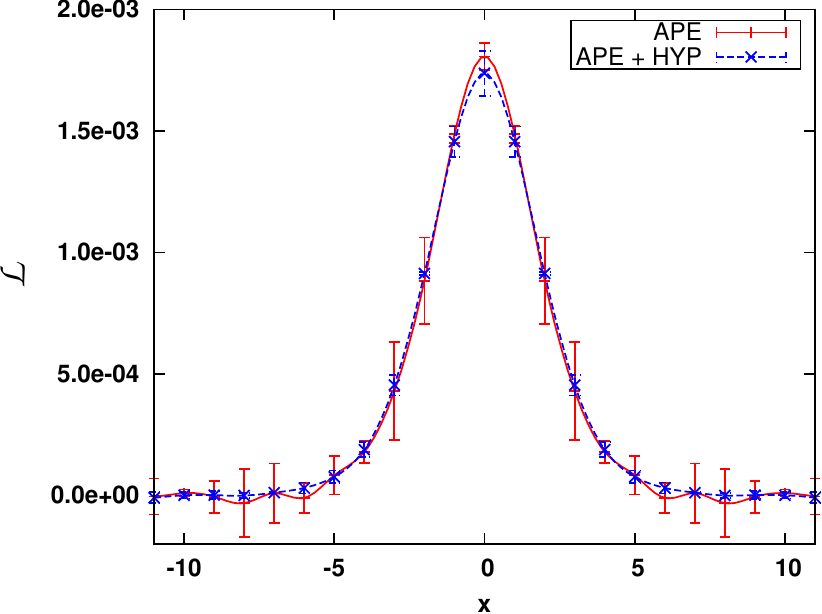}
    \caption{Comparison between results for lagrangian density with and without Hypercubic Blocking (HYP) smearing in time.}
    \label{APE_HYP}
\end{center}
\end{figure}

In this work two geometries for the hybrid system, $gq\overline{q}$, are investigated: a U shape and a L shape geometry, both defined in Fig. \ref{shape}. In the L shape geometry only the case when the gluon and the antiquark are superposed, the quark-antiquark case, is studied.
The use of the APE (in space) and HYP (in time) smearing allows us to have better results for the flux tube, Fig. \ref{APE_HYP}, while suppressing the fields near the sources.

\subsection{U profiles and Casimir scaling}

\begin{figure}[h]
\begin{centering}
    \subfloat[Chromoelectric Field\label{qqg_U_ape_hyp_E_y=4}]{
\begin{centering}
    \includegraphics[height=4cm]{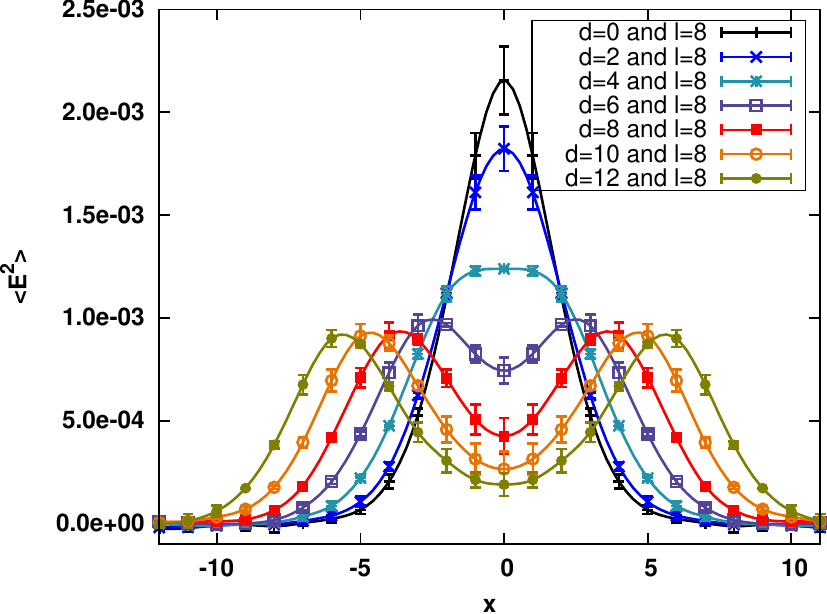}
\par\end{centering}}
    \subfloat[Chromomagnetic Field\label{qqg_U_ape_B_Act_y=4}]{
\begin{centering}
    \includegraphics[height=4cm]{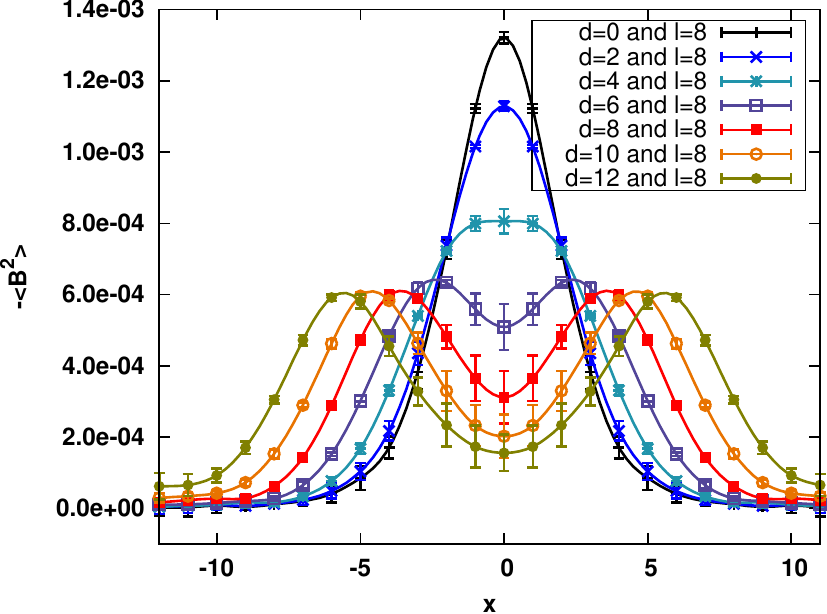}
\par\end{centering}}

    \subfloat[Lagrangian Density\label{qqg_U_ape_hyp_Act_y=4}]{
\begin{centering}
    \includegraphics[height=4cm]{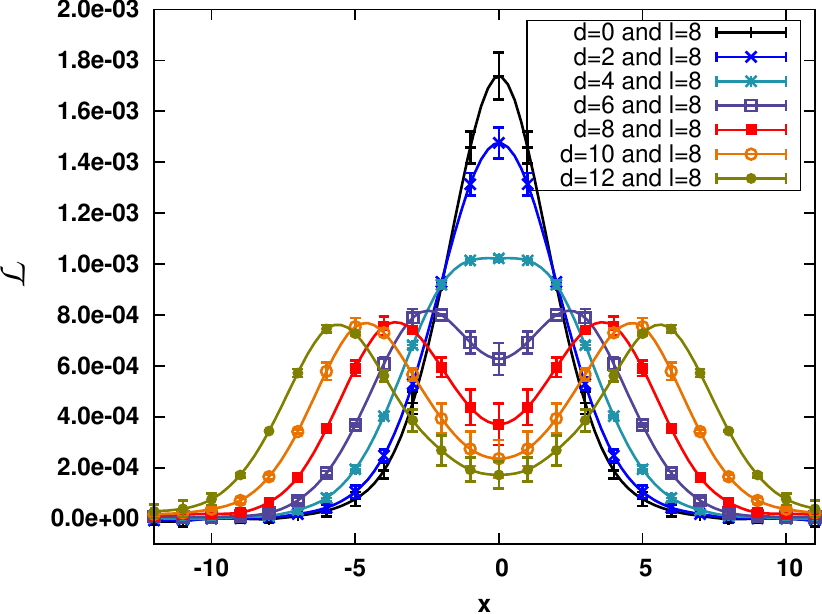}
\par\end{centering}}
    \subfloat[Energy Density\label{qqg_U_ape_hyp_Energ_y=4}]{
\begin{centering}
    \includegraphics[height=4cm]{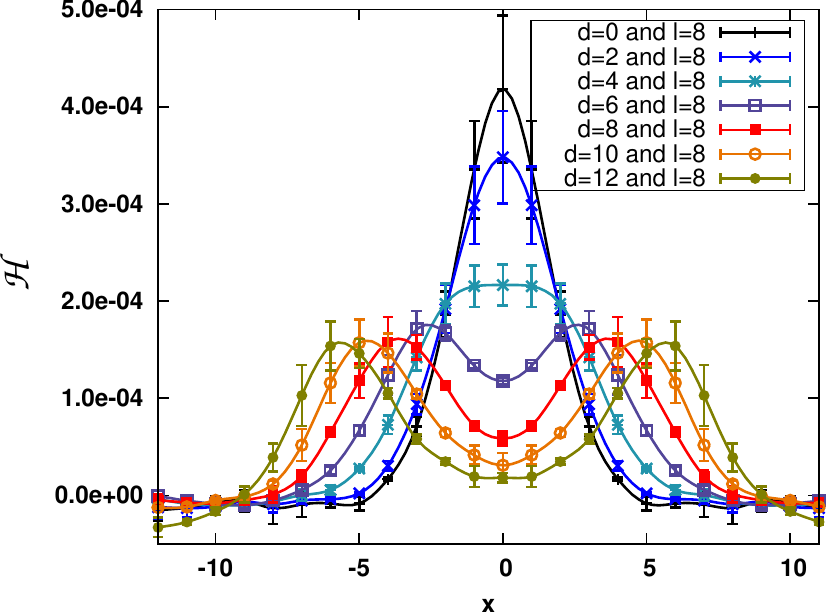}
\par\end{centering}}
\par\end{centering}
    \caption{Results for the U geometry at $y=4$ and $z=0$.}
    \label{qqg_U_Sim_profile}
\end{figure}

In Fig. \ref{qqg_U_Sim_profile} we present the profiles for the U geometry for $l=8$ and $d$ between 0 and 16 at $y=4$.
We can see the stretching and partial splitting of the flux tube in the equatorial plane ($y = 4$) between the quark and antiquark.
For $d=2$ and 4 at $y=4$ the separation between the two flux tube is not visible, this is due to the overlap in the tails of the flux tube which contributes for the total field, and for large separations the tails of the flux tubes contributes to a non zero field at $x=0$.

\begin{figure}[h]
\begin{centering}
    \subfloat[$r=y$ at $x=0$ and $z=0$\label{casimir_xy_Energ_x=0}]{
\begin{centering}
    \includegraphics[height=4.5cm]{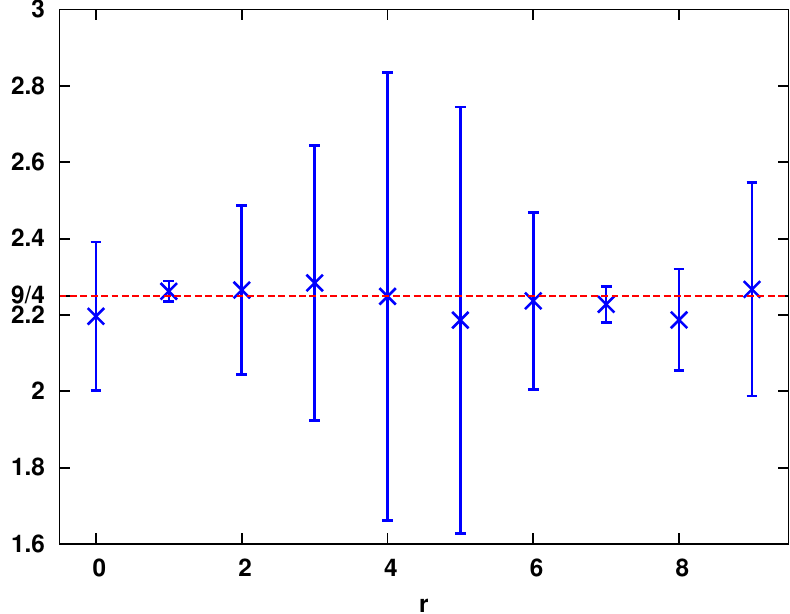}
\par\end{centering}}
    \subfloat[$r=(x,y)$ at $y=4$\label{casimir_xz_Energ}]{
\begin{centering}
    \includegraphics[height=4.5cm]{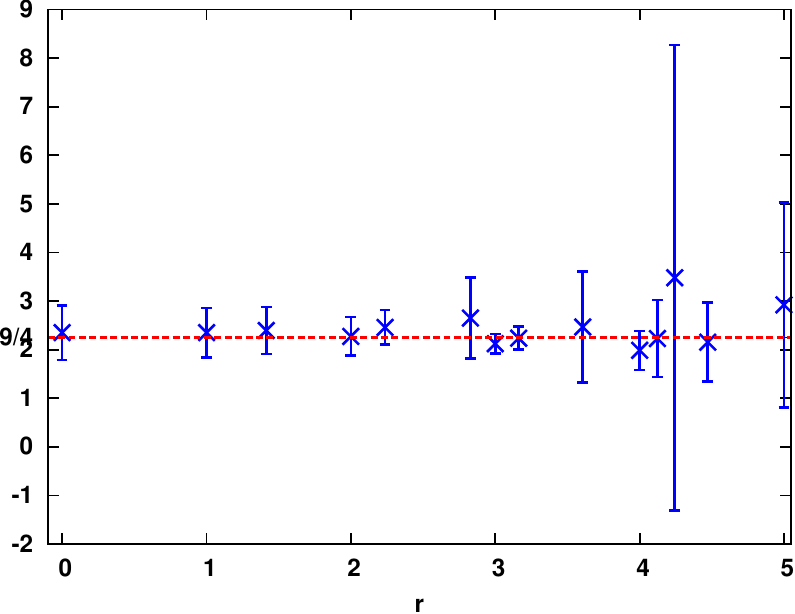}
\par\end{centering}}
\par\end{centering}
    \caption{Casimir scaling.}
    \label{casimir}
\end{figure}

We measure the quotient between the energy density of the two gluon glueball system and of the meson system, in the
mediatrix plane between the two particles ($x = 0$).
The results are shown in Fig. \ref{casimir}.
In Fig. \ref{casimir_xy_Energ_x=0} we present the results for $r=y$ at $x=z=0$ and in Fig. \ref{casimir_xz_Energ} we present the results for $r=(x,z)$ at $y=4$.
We make a constant fit to the data in Fig. \ref{casimir}, the result for Fig. \ref{casimir_xy_Energ_x=0} is $2.25096 \, \pm \, 0.0244972$ and for Fig. \ref{casimir_xz_Energ} is $ 2.23591 \, \pm \,0.0598732$.
As can be seen, these results are consistent with Casimir scaling, with a factor of $9/4$ between the energy density in the glueball and in the meson. This corresponds to the formation of an adjoint string.

\subsection{Dual Gluon Mass}

In 1970's, Nambu \cite{Nambu:1974zg}, 't Hooft \cite{'tHooft:1979uj} and Mandelstam \cite{Mandelstam:1974pi} proposed an interesting idea that quark confinement would be physically interpreted using the dual version of the superconductivity, the QCD vacuum state to behave like a magnetic superconductor.
The chromoelectric field originated by a $q\overline{q}$ pair is squeezed by Meissner
effect into a dual Abrikosov flux tube, giving rise to the confining linear potential, the field is confined into flux tubes, QCD strings. Color confinement could be understood as the dual Meissner effect.
In common superconductivity the magnetic field decays with $B \sim e^{-r/\lambda_L}$
and this could be interpreted in terms of an effective mass for the photon $m_\gamma = 1 / \lambda_L$.
There also is evidence for the dual superconductor picture from numerical simulations of QCD, some studies have point a similar behavior in QCD, \cite{Baker:1984qh,Bali:1996dm,Jia:2005sp}.

We tested two functions, $a\, e^{-2\mu r}$ and $a\, K_0^2 \left(\mu r\right)$, where $\mu=\frac{1}{\lambda_L}$, $\lambda_L$ is the penetration length and $K_0$ the modified Bessel function of order zero. So, in this case we have $\mu$ as the dual gluon mass.
Fitting the chromoelectric field and the lagrangian density section in the mid distance of the flux tube of the meson and the two gluon glueball, we obtain the results presented in Table \ref{tab_dual_gluon_mass} for the effective dual gluon mass, of the order of $ \sim 1\, \text{GeV} $.
Some values found in literature, for the effective dual gluon mass,
\cite{Suganuma:1997xk,Suganuma:2000jh,Suganuma:2002pm,Suganuma:2003ds}, and the effective gluon mass, \cite{Field:2001iu}, point for a same value.

\begin{table}[h]
\small{
\begin{centering}
\begin{tabular}{|c|c|c|c|c|}
\cline{2-5}
\multicolumn{1}{c|}{} & \multicolumn{2}{c|}{\T \B  $a\, e^{-2\, \mu\, r}$} & \multicolumn{2}{c|}{$a\, K_{0}^{2}\left(\mu\, r\right)$}\tabularnewline
\cline{2-5}
\multicolumn{1}{c|}{} & $\mu\ \left(\text{GeV}\right)$ &\T \B  $\chi^{2}/dof$ & $\mu\ \left(\text{GeV}\right)$ & $\chi^{2}/dof$\tabularnewline
\hline
\hline
\T \B $E_{(1)(a)}^{2}\left(r\right)$ & $1.170\pm0.228$ & $1.069$ & $0.805\pm0.287$ & $1.827$\tabularnewline
\hline
\T \B $\mathcal{L}_{(1)(a)}\left(r\right)$ & $1.170\pm0.119$ & $0.512$ & $0.865\pm0.188$ & $1.203$\tabularnewline
\hline
\T \B $E_{(2)(a)}^{2}\left(r\right)$ & $1.231\pm0.286$ & $1.547$ & $0.881\pm0.334$ & $2.084$\tabularnewline
\hline
\T \B $E_{(1)(b)}^{2}\left(r\right)$ & $1.210\pm0.056$ & $0.887$ & $0.897\pm0.085$ & $1.185$\tabularnewline
\hline
\T \B $\mathcal{L}_{(1)(b)}\left(r\right)$ & $1.208\pm0.068$ & $0.560$ & $0.909\pm0.099$ & $0.909$\tabularnewline
\hline
\T \B $E_{(2)(b)}^{2}\left(r\right)$ & $1.210\pm0.063$ & $1.162$ & $0.889\pm0.097$ & $1.262$\tabularnewline
\hline
\T \B $\mathcal{L}_{(2)(b)}\left(r\right)$ & $1.191\pm0.031$ & $1.066$ & $0.899\pm0.048$ & $1.106$\tabularnewline
\hline
\end{tabular}
\par\end{centering}
}
\caption{Results for the dual gluon mass, where (1) is for the two gluon glueball and (2) for the quark-antiquark cases, and (a) at $y=4$ and $z=0$ with $r=x$ and (b) at $y=4$ with $r=(x,z)$.}
\label{tab_dual_gluon_mass}
\end{table}

\section{Conclusions}
When the quark and the anti-quark are superposed, this corresponds to the formation of an
adjoint string between the two gluon and agrees with Casimir Scaling measured by Bali \cite{Bali:2000un}.
This can be interpreted with a type-II superconductor analogy for the confinement in QCD with repulsion of the fundamental strings and with the string tension of the first topological excitation of the string (the adjoint string) larger than the double of the fundamental string tension.

We present a value for the dual gluon mass of $\sim 1 \text{ GeV}$ which is gauge independent.


\begin{thebibliography}{99}
\bibitem{Bicudo:2007xp}
P.~Bicudo, et~al.,
Phys.Rev.D77:091504.

\bibitem{Cardoso:2007dc}
M.~Cardoso, et~al.,
PoS LAT2007 (2007) 293.

\bibitem{Cardoso:2009kz}
  M.~Cardoso, et~al.,
  Phys.\ Rev.\  D {\bf 81}, 034504 (2010).


\bibitem{Cardoso:2009qt}
  M.~Cardoso, et~al.,
  arXiv:0910.0133 [hep-lat].

\bibitem{Hasenfratz:2001hp}
A.~Hasenfratz, F.~Knechtli,
Phys. Rev. D 64-3 (2001) 034504.

\bibitem{MILC}
This work was in part based on the MILC collaboration's public lattice gauge
  theory code. \url{http://physics.indiana.edu/~sg/milc.html}.

\bibitem{Nambu:1974zg}
Y.~Nambu,
Phys. Rev. D10 (1974) 4262.

\bibitem{'tHooft:1979uj}
G.~'t~Hooft,
  Nucl. Phys. B153 (1979) 141.

\bibitem{Mandelstam:1974pi}
S.~Mandelstam,
  Phys. Rept. 23 (1976) 245--249.

\bibitem{Bali:1996dm}
G.~S. Bali, et al.,
Phys. Rev. D54 (1996) 2863--2875.

\bibitem{Baker:1984qh}
M.~Baker, et~al.,
Phys. Rev. D31 (1985) 2575.

\bibitem{Jia:2005sp}
  D.~Jia,
  arXiv:hep-th/0509030.

\bibitem{Suganuma:1997xk}
H.~Suganuma, et~al.,
Prog. Theor. Phys. Suppl. 131 (1998) 559--571.

\bibitem{Suganuma:2000jh}
H.~Suganuma, et~al.,
Nucl. Phys. A670 (2000) 40--47.

\bibitem{Suganuma:2002pm}
H.~Suganuma, et~al.,
  Nucl. Phys. Proc. Suppl. 106 (2002) 679--681.

\bibitem{Suganuma:2003ds}
H.~Suganuma, H.~Ichie,
Nucl. Phys. Proc. Suppl. 121 (2003) 316--319.

\bibitem{Field:2001iu}
J.~H. Field,
Phys. Rev. D 66~(1) (2002) 013013.

\bibitem{Bali:2000un}
G.~S. Bali,
Phys. Rev. D62 (2000) 114503.

\end{thebibliography}
\end{document}